# Optimal Tracking Current Control of Switched Reluctance Motor Drives Using Reinforcement Q-learning Scheduling

Hamad A. Alharkan, Sepehr Saadatmand, Mehdi Ferdowsi, Pourya Shamsi
Department of Electrical and Computer Engineering
Missouri University of Science and Technology
Rolla, Missouri, USA
haahx9@mst.edu, sszgz@mst.edu, ferdowsi@mst.edu, shamsip@mst.edu

*Abstract*— In this paper, a novel Q-learning scheduling method for the current controller of switched reluctance motor (SRM) drive is investigated. Q-learning algorithm is a class of reinforcement learning approaches that can find the best forward-in-time solution of a linear control problem. This paper will introduce a new scheduled-Q-learning algorithm that utilizes a table of Q-cores that lies on the nonlinear surface of a SRM model without involving any information about the model parameters to track the reference current trajectory by scheduling infinite horizon linear quadratic trackers (LQT) handled by Q-learning algorithms. Additionally, a linear interpolation algorithm is proposed to guide the transition of the LQT between trained Q-cores to ensure a smooth response as state variables evolve on the nonlinear surface of the model. Lastly, simulation and experimental results are provided to validate the effectiveness of the proposed control scheme.

*Index Terms*—Switched reluctance motor (SRM), Current control, Reinforcement learning (RL), Adaptive dynamic programming (ADP), Linear quadratic tracker (LQT), Least square (LS).

## I. INTRODUCTION

LATELY, switched reluctance motor (SRM) has earned significant consideration for a wide range of transportation electrification and variable speed applications . This is because it has several features such as a resilient and simple structure due to the lack of magnet, brushes, and rotor winding, and is efficient at high speed [1]. Based on the reduction in cost of power electronics, improved availability and performance of film capacitors to handle the pulse-type current of these machines, and the interest in reduced utilization of rare-earth magnets, the utilization of SRM for a variety of industrial and commercial applications has been on the rise [2]. This includes applications in traction drives as well as aeronautics where the high reliability, high temperature and vibration tolerance, and high-speed range of SRMs make them very competitive compared to more complex motors [3]-[7]. However, SRMs have suffered from certain drawbacks including high acoustic noise production due to its torque ripple and flux paths and high cost of drive due to the large number of semiconductor switches in its drive. Additionally, it has a high nonlinear electromagnetic nature that is highly reliant on variations in the phase current and rotor positions. Many researchers have investigated SRM to mitigate these issues by improving the SRM design to minimize torque ripples or developing a new converter topology using recently introduced and more affordable power electronic switches [8]-[11]. The high nonlinear behavior of SRM is the main challenge which must be considered when designing an effective controller.

Unlike conventional sinusoidal motors, SRM requires pulse type current that require high variations of current (i.e. di/dt) and hence a high bandwidth drive system. To achieve a fast rate of current charge and discharge, a large dc link voltage and low phase inductances are often needed. However, this dc link voltage will make the regulation of phase currents more challenging, particularly, during low speed operation modes. Traditionally, delta modulation or hysteresis current controllers have been used to regulate the phase current. Hysteresis-type controllers lead to a variable switching frequency which is not of interest as managing the Electro-Magnetic Interferences (EMI) becomes challenging. Additionally, power switches will impose an upper limit for the switching frequency and large current ripples will increase torque ripple and audible noise.

Many publications have investigated current control techniques for SRM including enhanced hysteresis control, sliding-mode approaches, and fast PI controllers [12]-[17]. However, PI-based methods are slow and methods such as delta-modulation won't be able to use the concept of duty-cycle to break-down the switching cycle to shorter active periods. To do so, a method to generate a duty cycle is needed. Classical control such as PID controllers are not capable of controlling a system with such transients. Hence researchers have investigated methods such as model predictive control and neural networks to cope with this issue [18]-[22]. To cope with nonlinearities of the model, [19] has introduced a Taylor expansion algorithm to approximate the variations of the model as a function of the rotor angle and current. Also, adaptive estimators are used to improve this approximation. However, the accuracy of the control is impacted by the Taylor expansion. As an improvement, [23] has introduced a table-based inductance function that is used to form the model needed for the MPC in each cycle in oppose to a Taylor expansion. This table allows the MPC to have access to an accurate inductance value for a given rotor angle and current. Additionally, an adaptive estimator is used to update this table. [23] has also introduced a linear interpolation technique for transitions between the models that will be incorporated in this paper to introduce a novel scheduled Q-learning technique. In these literatures, a fundamental model is assumed, then an adaptive estimator is used to estimate the inductance of the phase as function of rotor angle and current. Then, this value is used in a model predictive controller. The main drawbacks of above work are the need for a separate estimator, a model predictive controller, and assumptions on the structure of the model.

In this paper, the controller has been formulated based on an infinite-horizon linear quadratic tracker (LQT). To eliminate the need for a known model, a reinforcement Q-learning scheme is used to learn and apply the best course of action at each control cycle. Q-learning is inherently a linear controller [24] but the model of a SRM has nonlinearities to rotor angle and current (i.e. saturation). Hence, this paper proposes a scheduled Q-learning algorithm that utilizes a table of Q cores each containing a linear controller for a given rotor angle and current. By transitioning between these Q cores using a linear interpolation mechanism, this paper introduces a nonlinear tracking controller capable of handling SRM drives.

The specific contributions of this paper include i) introduction of Q-learning LQT for SRMs, ii) scheduling a table of Q-cores to achieve nonlinear control capabilities out of traditional Q-learning techniques, and iii) introducing a linear interpolation technique for transitioning between Q-cores to achieve a smooth Q scheduling.

The paper is organized as follows: Section II reviews the Q-learning algorithm and introduces the proposed controller. Section III proposes the Q scheduling algorithm and table interpolation. Sections IV and V verify the effectiveness of the proposed controller through simulations and experimental results.

## II. Q-LEARNING CONTROL OF SRM DRIVE

The primary target for the application of the Q-learning algorithm for the current control of SRM is solving the LQT problem which allows the system output to track a specific reference signal. The use of the algorithm in this way will minimize a predetermined value function associated with the cost of policy and the difference between the output current and reference signal. The classic solutions for LQT can be found by solving the feedback part using the algebraic Riccati equation (ARE) and a feedforward part using the noncausal difference equation [25]. However, these approaches are not applicable for the SRM or most industrial applications since they are solved offline and they need accurate information on system dynamics. Adaptive dynamic programming is part of the RL methods and has been used to solve infinite-horizon LQT problems online without knowing system dynamic [26]. Two major assumptions have been made for this scheme: full state feedback is observable for the controller and the full reference trajectory is known. LQT is a special case of model predictive controllers were the performance index is quadratic and no further constraints are applied to the optimizer. Deriving the quadratic form of the performance index for LQT has been proved in [26]. The benefit of quadratic forms is the availability of algorithms that can solve Bellman equations online. To cope with the reference trajectory, an augmented system is generated by incorporating the reference current trajectory into the state space model of SRM. This augmented system leads to the development of ARE which provides the optimal solution for LQT. By solving ARE, the feedback and feedforward parts of the policy for the classic solution of LQT are solved at the same time [27]. The main drawback of using the LQT Bellman equation to solve this problem is that the accurate model of SRM is needed [28].

To cope with this issue, Q-learning is utilized to learn and adapt to the optimal solution of this LQT online. The LQT Bellman equation and Q-learning algorithm for the SRM drive system are introduced in this section.

### A. LQT Bellman Equation Algorithm of SRM Drive

Driving the SRM requires a train of current pulses applied with respect to the rotor position. An assumption has been applied to this model to neglect the mutual inductance between the coils to minimize the system's complexity and decrease computational load. Taking into consideration the reference current, an augmented system can be formed by discretizing the SRM model using the forward approximation as

$$X_{k+1} = \begin{bmatrix} A & 0 \\ 0 & F \end{bmatrix} \begin{bmatrix} x_k \\ r_k \end{bmatrix} + \begin{bmatrix} B \\ 0 \end{bmatrix} u_k \equiv A_a X_k + B_b u_k$$

$$Y_k = [C \quad 0] x_k \equiv C_c X_k \qquad (1)$$

where $X_k = [x_k \ r_k]^T$ and $A = 1 - TR/L_k$, $B = T/L_k$, $x_k$ is the phase current, $u_k$ is the DC voltage bus, $R$ is the phase resistance, and the original output of the system $y_k$ is the phase current. $L_k$ is the nonlinear phase inductance as a function of both phase current and rotor position. $T$ is the sampling time and $F$ is the model of the reference trajectory (i.e. $F = 1$ for a flat current.) Due to the actual mechanical design of the machine, the nature of the inductance surface with respect to a rotor angle is periodic starting from an unaligned position between rotor and stator poles until they are aligned. To solve the LQT problem and achieve tracking, the reference current generator pulses are assumed to be incorporated in the augmented system as in (1). It is expressed as

$$r_{k+1} = F r_k \qquad (2)$$

where $r_k$ is the reference current trajectory and $F \in \mathbb{R}^n$ is the reference current generator. This can generate different types of waveforms including a sequence of square waveforms, the reference current for SRM. Even the command generator is not stable, then solving the LQT problem can occur by injecting the discount factor into the value function. Based on the augmented system, the discounted value function can be expressed as

$$V(x_k) = \frac{1}{2} \sum_{i=k}^{\infty} \gamma^{i-k} [X_i^T Q_q X_i + u_i^T R u_i] \qquad (3)$$

Where $Q_q = [C \ -I]^T Q [C \ -I]$, $Q$ and $R$ are predefined weight matrices for the augmented state and the control input, respectively, and $0 < \gamma \leq 1$ is a discount factor. The value of $\gamma$ should be less than 1 to attain a stable value function as the reference current in SRM is generated as a train of pulses and therefore has a positive dc average [29]. Based on (4), the value function relies on the current augmented state and an infinite horizon of the control inputs. By initializing the state of the value function with a fixed control input, that infinite sum can be written as

$$V(x_k) = \frac{1}{2}[(r_k - y_k)^T Q(r_k - y_k) + u_k^T R u_k] + \gamma V(X_{k+1}) \quad (4)$$

This formula is equivalent to LQT bellman equation. As it has been proved in [26] that the value function can be derived in a quadratic form and $V(x_k) = \frac{1}{2}X_k^T P X_k$, the LQT Bellman equation with respect to a kernel P matrix is generated as

$$X_k^T P X_k = x_i^T Q_q x_i + u_k^T R u_k + \gamma X_{k+1}^T P X_{k+1} \quad (5)$$

Where $P$ matrix is the optimum solution of ARE with elements derived in [26]. By obtaining the Hamiltonian function of LQT and applying the stationary condition to obtain the optimal control policy (8), the solution of ARE that allows matrix $P$ to converge to its optimal values can be generated as

$$P = Q_q + \gamma A_a^T P A_a - \gamma^2 A_a^T P B_b (R + \gamma B_b^T P B_b)^{-1} B_b^T P A_a \quad (6)$$

Now, one may construct the algorithm based on the policy iteration method to solve the LQT problem by iterating the Bellman equation until convergence using data measured during the operation of the machine as in algorithm 1 as follows.

### B. Q-learning Algorithm of SRM Drive

Let's assume that $L_k$ and hence the model of the machine is linear. For instance, the controller is operating while the variations of the current and angle of the rotor are negligible. This is due to the fact that the Q-learning algorithm utilized in this section can only operate on linear systems. In the next section, the nonlinearity is addressed through scheduling.

In Algorithm 1, Policy Integration (PI) is applied to LQT Bellman equation to acquire the optimum solution for ARE.

---

**Algorithm 1**: Solving LQT Bellman equation online by using PI
**Initialization**: Initialize the algorithm with stable control input. Repeat and update the following two process until convergence.
1) **Policy Evaluation:**
$$X_k^T P^{i+1} X_k = (X_k^T)(Q_q + (K_P^i)^T R(K_P^i))(X_k) + \gamma X_{k+1}^T P^{i+1} X_{k+1} \quad (7)$$

2) **Policy Improvement:**
$$K_P^{i+1} = (R + \gamma B_b^T P B_b)^{-1} \gamma B_b^T P A_a \quad (8)$$

---

This algorithm requires all SRM dynamic parameters (i.e. $A_a$) to solve the LQT problem online. Q-learning is among the RL control methods that offer an adaptive tuning algorithm to track the reference signal online without requiring the system dynamic [30]. By extracting sets of data during the operation, including the reference current and augmented states, the algorithm can train Q-function until convergence at each iteration. The Q-function of LQT can be provided in matrix form by substituting the augmented model (1) and reference current in the LQT Bellman equation as

$$Q(X_k, u_k) = \frac{1}{2}\begin{bmatrix}X_k\\u_k\end{bmatrix}^T \begin{bmatrix}Q_q + \gamma A_a^T P A_a & \gamma A_a^T P B_b\\ \gamma B_b^T P A_a & R + \gamma B_b^T P B_b\end{bmatrix}\begin{bmatrix}X_k\\u_k\end{bmatrix} \quad (9)$$

which can be written as

$$Q(X_k, u_k) = \frac{1}{2}\begin{bmatrix}X_k\\u_k\end{bmatrix}^T \begin{bmatrix}G_{XX} & G_{Xu}\\ G_{uX} & G_{uu}\end{bmatrix}\begin{bmatrix}X_k\\u_k\end{bmatrix} \quad (10)$$

The Q-learning algorithm can be designed based on the Policy iteration method to solve LQT online in a way that ensures the system model parameters do not appear in the algorithm processes [31]. This process improves the control input until the system converges to the optimal level which allows the output current in SRM to follow the reference current. The Q-learning Algorithm 2 to find the solution of ARE is as follows

---

**Algorithm 2**: Solving LQT Q-Function online by using PI
**Initialization**: Initialize the algorithm with stable control input. Repeat and update the following two process until convergence.
1) **Policy Evaluation:**
$$M_k^T G^{i+1} M_k = (X_k^T)Q_q(X_k) + (u_k^i)^T R(u_k)^i + \gamma M_{k+1}^T G^{i+1} M_{k+1} \quad (11)$$

2) **Policy Improvement:**
$$u_k^{i+1} = -(G_{uu}^{-1})^{i+1} G_{uX}^{i+1} X_k \quad (12)$$

---

Where $M$ is defined as $M = [X_k \quad u_k]^T$. Optimizing the Q-function in Algorithm 2 can be achieved as G matrix trains and converges to the optimum solution. The policy evaluation step for both algorithm 1 and 2 requires solver to achieve convergence before updating the policy [28].

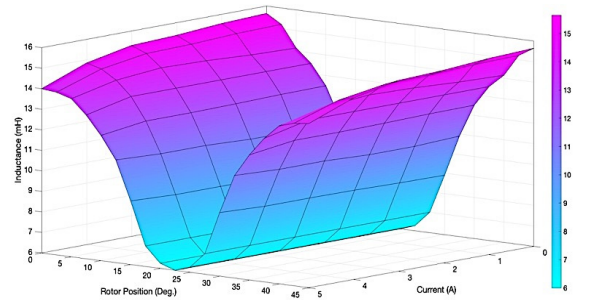

Fig 1. The inductance profile with respect to the rotor angles and the currents.

## III. Q-LEARNING SCHEDULING

In previous section, the adaptive Q-learning algorithm controller for SRM was proposed to solve LQT and enable the current of the SRM drive to track a reference trajectory assuming that $L_k$ was constant. The inductance profile of SRM is a nonlinear function of current and the rotor angle. For instance, the inductance profile of the motor utilized later in the experimental section, is shown in fig. 1. In addition to this function, in long term, effects such as aging of bearings and changes in the airgap, chemical degradation of the core such as rust which can lead to changes in the airgap length, and temperature expansion can cause further variation in the inductance profile. Also, common manufacturing related variations such as variations in the airgap length, permeability, and even number of turns can cause some differences between the expected model and the actual inductance profile. To mitigate these effects, adaptive estimation approaches to update the dynamic parameters of the machine are of interest. Various methods have been utilized to estimate the inductance profile of SRM and update the nonlinear model of the SRM [32]-[33]. However, these methods are unlike the proposed Q-learning approach which can perform both tracking reference and adaptive estimation in the same time. Q-learning by itself is not feasible or applicable to a nonlinear system such as a SRM. To address this issue, one can incorporate a proper local linearization scheme for the nonlinear inductance surface of SRM to allow the Q matrix to train in its locally linearized region.

Gain scheduling is a powerful solution to enable a linear control solution to address a nonlinear control problem. Gain scheduling is commonly applied to classical PID controllers to fine tune control parameters for the local operating conditions. Several implementation of gain scheduling methods have been studied for SRM control such as in speed control [34] and in PI current controller for enhancing the performance [35]. Gain scheduling allows the Q-learning algorithm to react rapidly to variations in operating conditions. For this, it is important to select enough Q-cores to reflect the nonlinear system properly. In contrast to nonlinear RL methods for adaptive dynamic programming such as neural networks which require heavy matrix operations for online training, gain scheduling provides less computation loads which makes it ideal for implementations on conventional micro-controllers. Its major requirement is access to sufficient memory to store all trained Q matrix in the table elements based on the corresponding states and rotor angles. The data table must be updated for each iteration during system trajectories.

In this paper, the surface of the inductance profile of SRM is divided into enough segments to achieve a suitable linearization with a balanced tradeoff between the number of table elements and accuracy. Each Q-core represents a local linear controller that is capable of following the trajectory for that specific region of operation. The Q-learning algorithm can be executed in each segment by training a Q matrix in the segment until the matrix achieves its optimum solution based on the data collected for the segment. The training Q matrix at each segment is registered and stored as a table entry to generate a bidimensional Q matrix array which helps to efficiently conquer the nonlinear behavior of SRM.

For a fast control response, the Q-core table is fully trained and solved offline using expected motor parameters and the proposed algorithm and then preloaded into the control system. This allows the controller to only adapt to the variations between the expected and actual model. The following subsections show the process of the Q-scheduling algorithm and its stages. The first stage involves solving and training Q-functions at each Q-core and can be performed using the least square method. Then, table implementation method to transmit the scheduled Q matrix from the Q-table to the policy improvement to update the control input.

### A. Training Local Q Matrices

In this paper, the least square approach has been utilized to solve the tracking problem and learn the Q matrix by using enough data packets measured through the operating of the machine. The least square solver does not require system model identification model. In practice, an observer is required to observe that states online. To implement policy evaluation, no less than $H = (m_x + m_u + m_y) \times (m_x + m_u + m_y + 1)/2$ data tuples are needed to perform LS method while $Q(X_k, u_k) = \frac{1}{2} M_k^T G M_k$ and the number of elements in G matrix are $(m_x + m_u + m_y) \times (m_x + m_u + m_y)$. This can be solved using the Kronecker product which enables the Q matrix to appear as a columns of stacking vectors as

$$\mathcal{A}(vec(G)^T) = \mathcal{B} \tag{13}$$

The definition of $\mathcal{A}_k$ and $\mathcal{B}_k$ are expressed as

$$\mathcal{A} = \begin{bmatrix} M_k \otimes M_k - \gamma M_{k+1} \otimes M_{k+1} \\ \vdots \\ M_{k+z} \otimes M_{k+z} - \gamma M_{k+z+1} \otimes M_{k+z+1} \end{bmatrix} \tag{14}$$

and

$$\mathcal{B} = \begin{bmatrix} (X_k^T) Q_q (X_k) + (u_k^i)^T R (u_k)^i \\ \vdots \\ (X_{k+z}^T) Q_q (X_{k+z}) + (u_{k+z}^i)^T R (u_{k+z})^i \end{bmatrix} \tag{15}$$

where $z \geq H$ is the number of samples for each iteration. Then, the batch least square equation for solving Q matrix is provided as

$$vec(G^{j+1}) = (\mathcal{A}^T \mathcal{A})^{-1} \mathcal{A}^T \mathcal{B} \tag{16}$$

By maintaining the persistence condition, least square may be solved iteratively by applying recursive least square (RLS) equations as

$$e_k(t) = Q(X_k, u_k) - \mathcal{A}_k^T \overline{G_k}(t-1)$$

$$\overline{G_k}(t) = \overline{G_k}(t-1) + \frac{\eta_k(t-1) \mathcal{A}_k e_k}{1 + \mathcal{A}_k^T \eta_k(t-1) \mathcal{A}_k}$$

$$\eta_k(t) = \eta_k(t-1) - \frac{\eta_k(t-1) \mathcal{A}_k \mathcal{A}_k^T \eta_k(t-1)}{1 + \mathcal{A}_k^T \eta_k(t-1) \mathcal{A}_k} \tag{17}$$

where $t$ is the index of iterations of the RLS, $e$ is the error, and $\eta$ is the covariance matrix whereas $\eta_k(0) = \tau I$ for a big positive number $\tau$ while $I$ is an identity matrix.

## B. Table Data Extraction and Linear Interpolation

Table readout algorithm is important to enable extracting the knowledge from the Q-cores table and utilizing the data to improve policy. A table of Q-learning has been computed and formed in the previous section that contains the locations for current-rotor position points selected from the surface of the inductance profile. The typical current pulse for each SRM phase placed on the Q-learning table is shown in Fig. 2a. One method of implementing the Q-learning table is to use the optimal Q matrix that is located at the near current path. In this case, the algorithm will read the value of the current and measure the distance to neighboring matrices to find the nearest Q matrix. This process solves the problem of using only one learned Q matrix in the locally linearized region. However, in practice, this method is pretty simple. However, it leads to transients in the current waveform every time the controller switches between two table elements.

The bilinear interpolation algorithm provides a smoother and a more accurate scheduling than does the nearest Q matrix method. This algorithm divides the Q matrix among its four closest Q matrices neighbors in the opposite proportion of the distance, which means if the state of the system is located at equal distance from four Q neighbors, the values of scheduled Q matrix are divided equally; if it is a near one of the four matrices, most of the scheduled Q matrix data are transmitted from that adjacent Q matrix. Observing the four neighboring Q matrices points $Q_{11}$, $Q_{12}$, $Q_{21}$ and $Q_{22}$, which are the four closest neighbors of scheduled Q matrix $Q_s$, then $Q_s$ is obtained as

$$Q_s = \beta_0 + \beta_1 \theta + \beta_2 i + \beta_3 \theta i \quad (18)$$

where the coefficient of bilinear scheduling $\beta_0$, $\beta_1$, $\beta_2$ and $\beta_3$ are obtained by solving

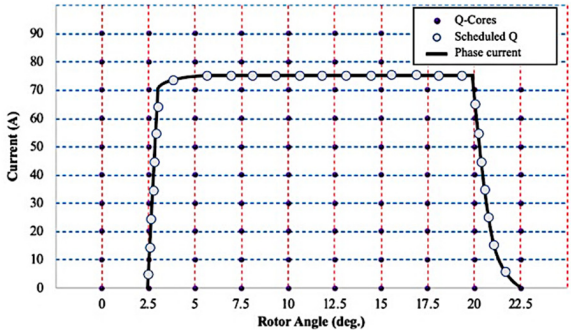

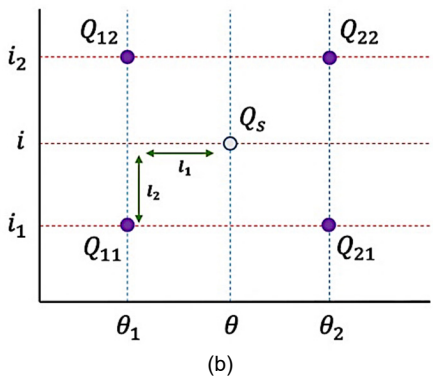

(a)

(b)

Fig 2. The process of implanting the Q-cores table into the controller, (a) The sample current path lies on the Q-cores table, (b) the definition of the bilinear scheduling parameters.

$$\begin{bmatrix} \beta_0 \\ \beta_1 \\ \beta_2 \\ \beta_3 \end{bmatrix} = \begin{bmatrix} 1 & \theta_1 & i_1 & \theta_1 i_1 \\ 1 & \theta_1 & i_2 & \theta_1 i_2 \\ 1 & \theta_2 & i_1 & \theta_2 i_1 \\ 1 & \theta_2 & i_2 & \theta_2 i_2 \end{bmatrix}^{-1} \begin{bmatrix} Q_{11} \\ Q_{21} \\ Q_{12} \\ Q_{22} \end{bmatrix} \quad (19)$$

In practical implementation, to avoid solving systems of equations and performing matrix inversions that are not feasible in a digital controller, and since the scheduled Q matrix lies on a square grid of four Q matrices, one can use a simplified algorithm based on a unit square, $Q_s$ is computed as

$$Q_s = [1 - l_2 \quad l_2] \begin{bmatrix} Q_{11} & Q_{12} \\ Q_{21} & Q_{22} \end{bmatrix} \begin{bmatrix} 1 - l_1 \\ l_1 \end{bmatrix} \quad (20)$$

where $l_1 \in [0,1)$ and $l_2 \in [0,1)$ are the lengths between $Q_s$ and the nearest Q matrix in the rotor angels and current axis, respectively. These lengths are calculated as shown in Fig. 2b as $l_1 = (\theta - \theta_1/\theta_2 - \theta_1)$ and $l_2 = (i - i_1/i_2 - i_1)$. Implementing this method drastically minimizes the computational burden of the scheduling process and the number of cycles required for scheduling.

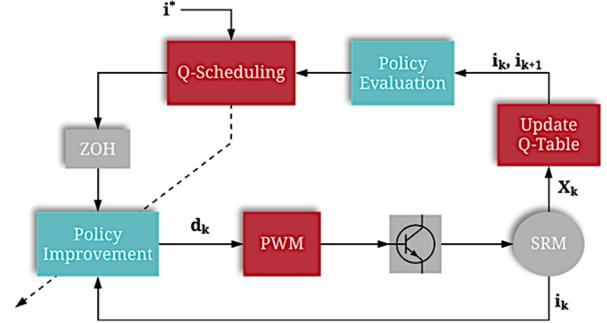

Fig 3. The overall control scheme.

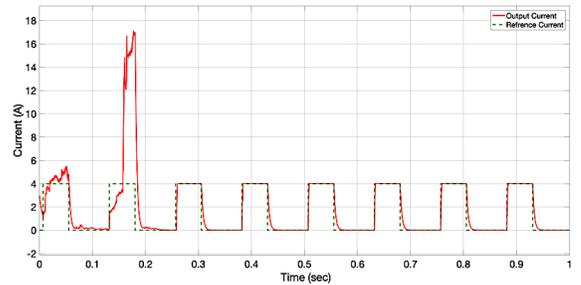

Fig 4. Tracking the output current of SRM to the reference current.

## IV. SIMULATION RESULT

The Q-learning algorithm integrated with the bilinear scheduling approach has been simulated to study the performance of the proposed current controller and verify the effectiveness of the controller. The control scheme is depicted in Fig. 3. This controller has been applied to a 500 W 12/8 SRM, which has a phase resistance of 2 Ω and a nominal current of 5 A. The inductance profile of the controller begins from the aligned position at 16 mH and gradually decreases until it reaches the unaligned position at 6 mH. The simulation sampling time is at 0.1 ms. Algorithm 2 has been utilized to train all Q-cores pre-located on the nonlinear surface of the machine. In this case, the algorithm should initialize the process using a stable control policy and augmented state. The initial augmented state and initial control policy have been selected to be $X_0 = [0 \ \ 0]^T$ and $K_0 = [100 \ \ -100]^T$, respectively. The cost function has been applied with the weights of Q = 100 and R = 0.001. The discount factor in this function is $\gamma = 0.9$. The reference model generates a train of square wave signals, the typical reference current for SRM. The Q-matrix at the unaligned position and a current of 4 A converges to its optimal values to allow tracking performance as shown below:

$$G = \begin{bmatrix} 438100 & -253100 & 5729 \\ -253100 & 56630 & -2595 \\ 5729 & -2595 & 98.1 \end{bmatrix} \quad (21)$$

And, the optimal control gain K converges to

$$K = [120 \ \ -122] \quad (22)$$

The optimal values vary based on Q-core along with the states that are located in the domain of the system. For each Q-core, there are 6 data tubules collected per iteration to train the Q-matrices using the LS square method. In this simulation, the speed of the SRM is constant and has been selected to be 60 RPM to demonstrate the result for the proposed controller. Fig. 4 shows how the SRM drive current tracks the reference of sequent pulses within a few time steps. Fig. 5 shows how the control gains K values that have converged to their optimal numbers change (considering the movement along the scheduling-table as well). The optimal voltage signal introduced to the motor to verify the best tracking performance is shown in Fig. 6. Fig. 7 shows the behavior of the current once the reference changed from 4A to 5.5A. This figure illustrates that any change in the reference current will allow the Q-table to start re-learning to adapt to that change.

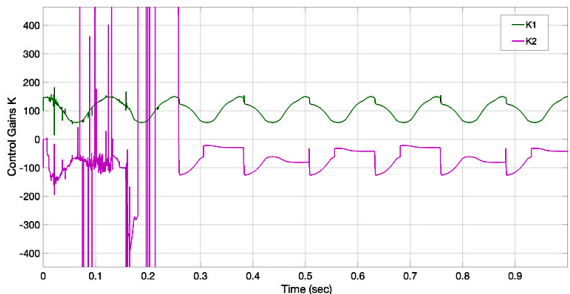

Fig 5. The convergence of the gains K values throughout learning process.

## V. EXPERIMENTAL RESULTS

In this section, the results and observations are presented to show the practical feasibility of the control method. The experimental components include a 3-phase 500 W 12/8 SRM, DC machine with DC power supply to control the field and hence loading of the machine as a mechanical load, H bridge converter, control board with a TI TMS320F28377D microcontroller, and a mixed domain oscilloscope. The experimental setup is shown in Fig. 8. The unaligned and aligned inductances for the machine used for validation are 6mH and 16mH, respectively, and the nominal current is 5 A per phase. The proposed Q-learning algorithm is implemented inside one of the TMS320F28377D cores capable of operating at 200MFLOPS. Learning Q-matrices along with the updating of the Q-table and the scheduling mechanism processes require approximately 6,000 cycles which is less than the 10,000 cycles available between 2 switching cycles. This shows that the controller is feasible for the proposed algorithm and hence the proposed algorithm is suitable for industrial/commercial deployment.

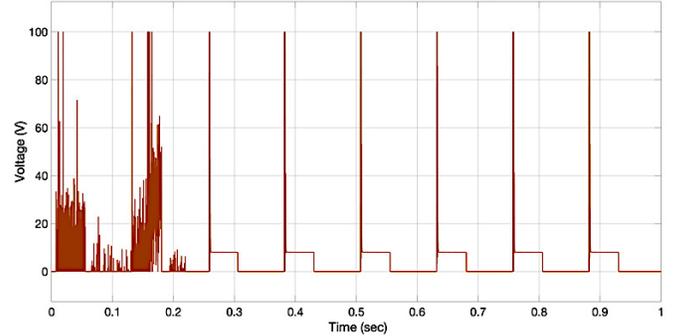

Fig 6. The optimal phase voltage introduced to the machine throughout learning.

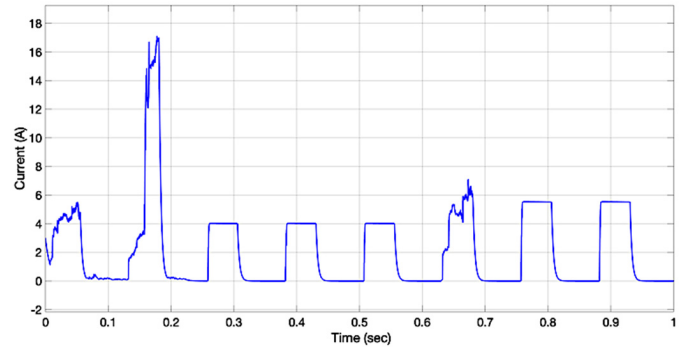

Fig 7. The behavior of the current when the reference changed.

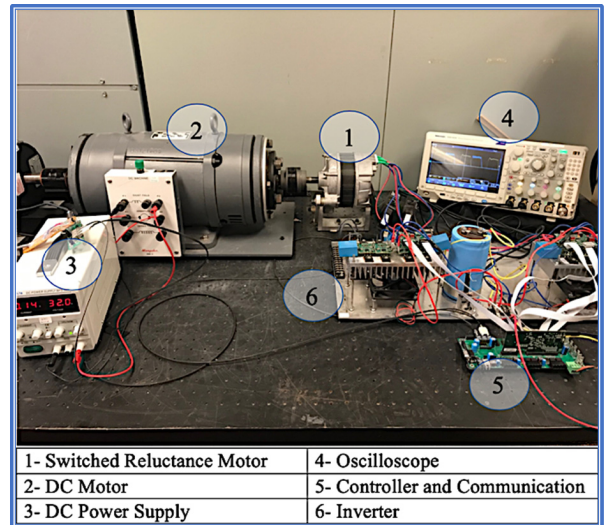

Fig 8. The experimental setup.

### A. Tracking Performance and Comparisons

The behavior of the current at different stages of the learning process is shown in Fig. 9. In this figure, the controller is set to run starting from the preloaded Q table to the point that the Q table is trained to the actual hardware online. In this figure, probe 1 shows the behavior of Phase A under the proposed control while probe 2 shows Phase B under the traditional Delta-modulation for comparison. During the learning process when the Q matrices are not fully trained, the current tries to track the reference current (Fig. 9a). The zoomed version of the current response is shown in Fig. 9b.

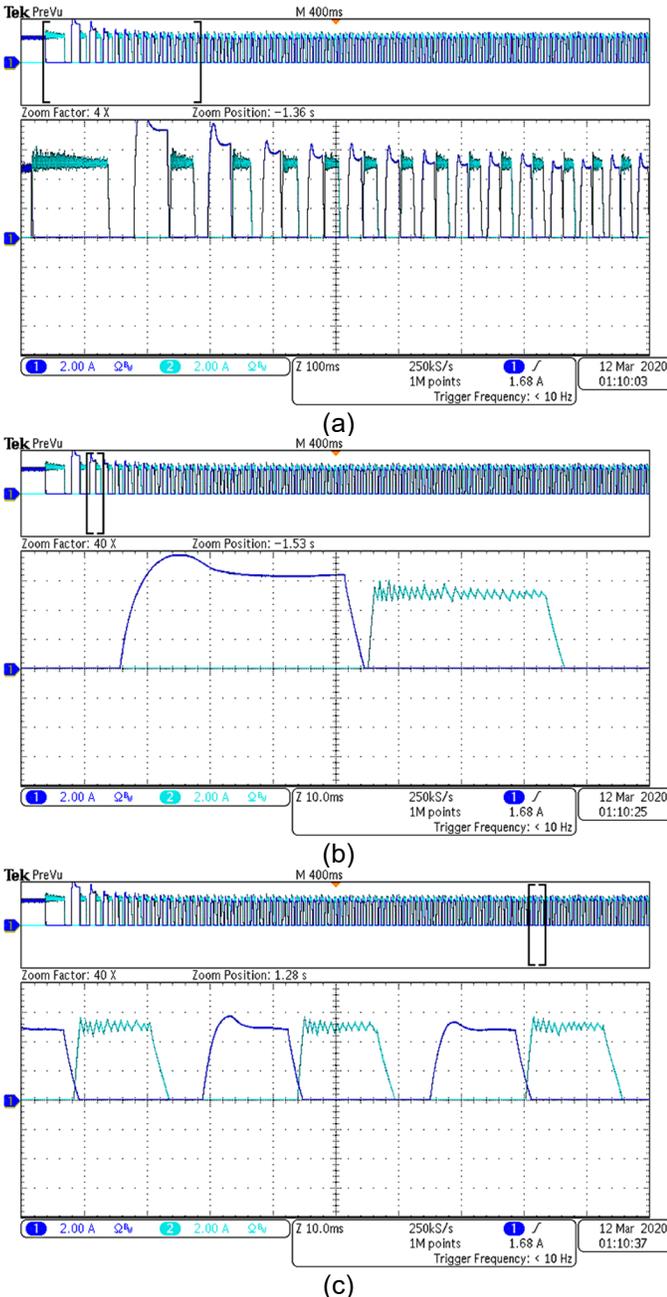

Fig 9. Start up process with the current set on 5.5 A (a) when the Q matrices are not fully trained, (b) the zoomed version of current response when the Q matrices are not fully trained (c) the current response when the Q matrix is fully trained.

After a couple of cycles, the Q matrices are fully trained and the current can successfully track the reference current with almost no ripples on the current pulses (Fig. 9c). Delta-modulation is not effective in minimizing the ripples for the current pulses. The Q-learning algorithm, once the Q-matrices are fully trained, are much more effective at minimizing the ripples for current pulses.

### B. Changing the Reference Current Response

In this test, the reference current is changed from 5.5 A to 4.5 A. The observations are illustrated in Fig. 10. This figure shows that the Q-matrices begin retraining once the reference current varies to 4.5 A. Fig. 10a shows how the new reference is tracked. After a few cycles, the current tracks the reference effectively after the Q-matrices are fully trained (Fig. 10b). When conventional delta-modulation is used, large ripples will be observed in phase current and there is no way to mitigate them (Fig. 10b).

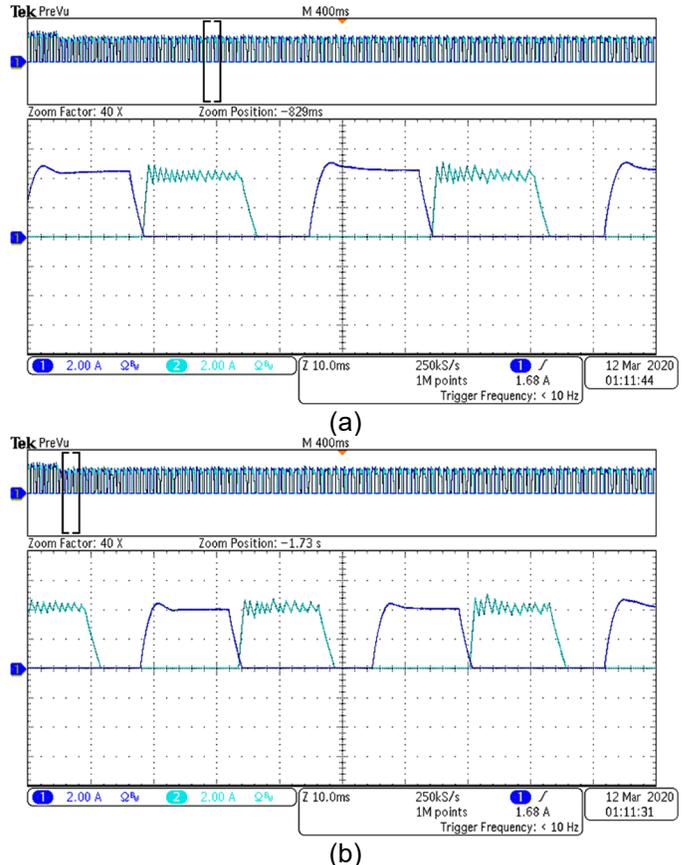

Fig 10. Reference change from 5.5 to 4.5A (a) when the Q matrices are not fully trained, (b) When the Q matrices are trained.

## VI. CONCLUSION

The Q-learning scheduling algorithm for controlling the current of a SRM drive was studied in this paper. After the introduction a Q-learning LQT, a table of Q-cores was generated to cover the nonlinear surface of SRM's model. Using this table, a scheduled Q-learning controller was derived which is capable of controlling a nonlinear system, particularly, a SRM drive. Additionally, an online training mechanism was introduced capable of controlling the SRM without having any information regarding its model parameters. This training mechanism updates each Q-core in the table as the state variables evolve over the domain of this table. Furthermore, a linear interpolation technique was used to ensure smooth transitions between these Q-cores. Lastly, simulation and experimental results demonstrated that the proposed algorithm is successful in controlling the current of a switched reluctance motor, minimizing its ripples, and adapting to the underlying SRM without any prior information regarding its parameters.